\documentclass[graybox]{svmult}

\usepackage{mathptmx}       
\usepackage{helvet}         
\usepackage{courier}        
\usepackage{type1cm}        
\usepackage{makeidx}         
\usepackage{graphicx}        
\usepackage{multicol}        
\usepackage[bottom]{footmisc}
\usepackage{amsmath} 
\usepackage{psfig}  

\makeindex             

\def\aap{A\&A}
\def\mnras{MNRAS}

\begin{document}

\title*{Pulsar Timing and its application for Navigation and Gravitational Wave Detection}
\titlerunning{Pulsar Timing, Spacecraft Navigation, Gravitational Wave Detection} 
\author{Werner Becker, Michael Kramer and Alberto Sesana}
\authorrunning{Becker, Kramer \& Sesana}
\institute{Werner Becker \at MPE, Garching, Germany \email{web@mpe.mpg.de}
\and Michael Kramer \at MPIfR, Bonn, Germany \email{mkramer@mpifr-bonn.mpg.de}
\and Alberto Sesana \at University of Birmingham, Birmingham, UK \email{asesana@star.sr.bham.ac.uk}}
%
%
\maketitle

\abstract{Pulsars are natural cosmic clocks. On long timescales they rival the precision of 
 terrestrial atomic clocks. Using a technique called pulsar timing, the exact measurement of 
 pulse arrival times allows a number of applications, ranging from testing theories of gravity 
 to detecting gravitational waves. Also an external reference system suitable for autonomous 
 space navigation can be defined by pulsars, using them as natural navigation beacons, not 
 unlike the use of GPS satellites for navigation on Earth. By comparing pulse arrival times 
 measured on-board a spacecraft with predicted pulse arrivals at a reference location (e.g.~the 
 solar system barycenter), the spacecraft position can be determined autonomously and with high 
 accuracy everywhere in the solar system and beyond. We describe the unique properties of    
 pulsars that suggest that such a navigation system will certainly have its application
 in future astronautics. We also describe the on-going experiments to use the clock-like 
 nature of pulsars to ``construct'' a galactic-sized gravitational wave detector for 
 low-frequency ($f_{GW}\sim 10^{-9} - 10^{-7}$ Hz) gravitational waves. We present the 
 current status and provide an outlook for the future.}

\newcommand{\klesssim}{\mathrel{\hbox{\rlap{\hbox{\lower4pt\hbox{$\sim$}}}\hbox{$<$}}}}

\section{Introduction}
\label{intro}

For millenia, keeping time was safely in the hands of astronomers, who
watched the heavens to serve the societies' needs for time
measurements.  They used the Earth's rotation, the moon, the Sun and
the stars to keep time, but, of course, they also exploited their
observations to study and explore the Universe that we live in.
Practical purposes, especially the need to navigate, turned the art of
timekeeping into the hands of clockmakers and, hence, engineers and
physicists. Today, world time is kept by a set of ultra-precise atomic
clocks, and the sophistication and accuracy of these clocks is hugely
impressive (see other contributions to this book).  The stability of
these clocks is best on small timescales. When time stability is
needed on long timescales, one can (and does) ``hand over'' time from
clock to clock -- or one can revert back to astronomical
observations. Moreover, the fact that the Earth is moving on an
elliptical orbit around the Sun, and hence at varying distances, means
that every clock on Earth experiences a varying gravitational
potential throughout the year. This leads to seasonal changes in clock
rates that affect all clocks on Earth simultaneously. Astronomy can,
hence, provide an ``independent'' time standard that is unaffected by
effects on Earth or in the solar system. The key in providing such an
astronomical time standard are objects called ``pulsars.''. Pulsars are 
compact, highly-magnetized rotating neutron stars which act as ``cosmic 
lighthouses'' as they rotate, enabling a number of applications as 
precision tools. 

This contribution describes pulsars, the technique of pulsar timing and 
some of the resulting applications. Coming full circle, these applications 
include navigation and also time keeping. Here we concentrate on the former
(see Section~\ref{sec:2}) and applications in fundamental physics,
especially the detection of low-frequency gravitational waves (see
Section~\ref{sec:3}). We note, that at the end, despite attractive
features of a pulsar timescale, pulsars will not be able to compete
with the precision and practicality of the best atomic clocks on
Earth. Nevertheless, it is the combination of pulsar clocks with
terrestrial clocks that allow, via clock comparison experiments, to
probe a wide range of physics -- and the Universe that we live in,
continuing the tradition of astronomers.

  \begin{figure}[h!!!!!!!!!!]
  \centerline{\includegraphics[width=10.5cm]{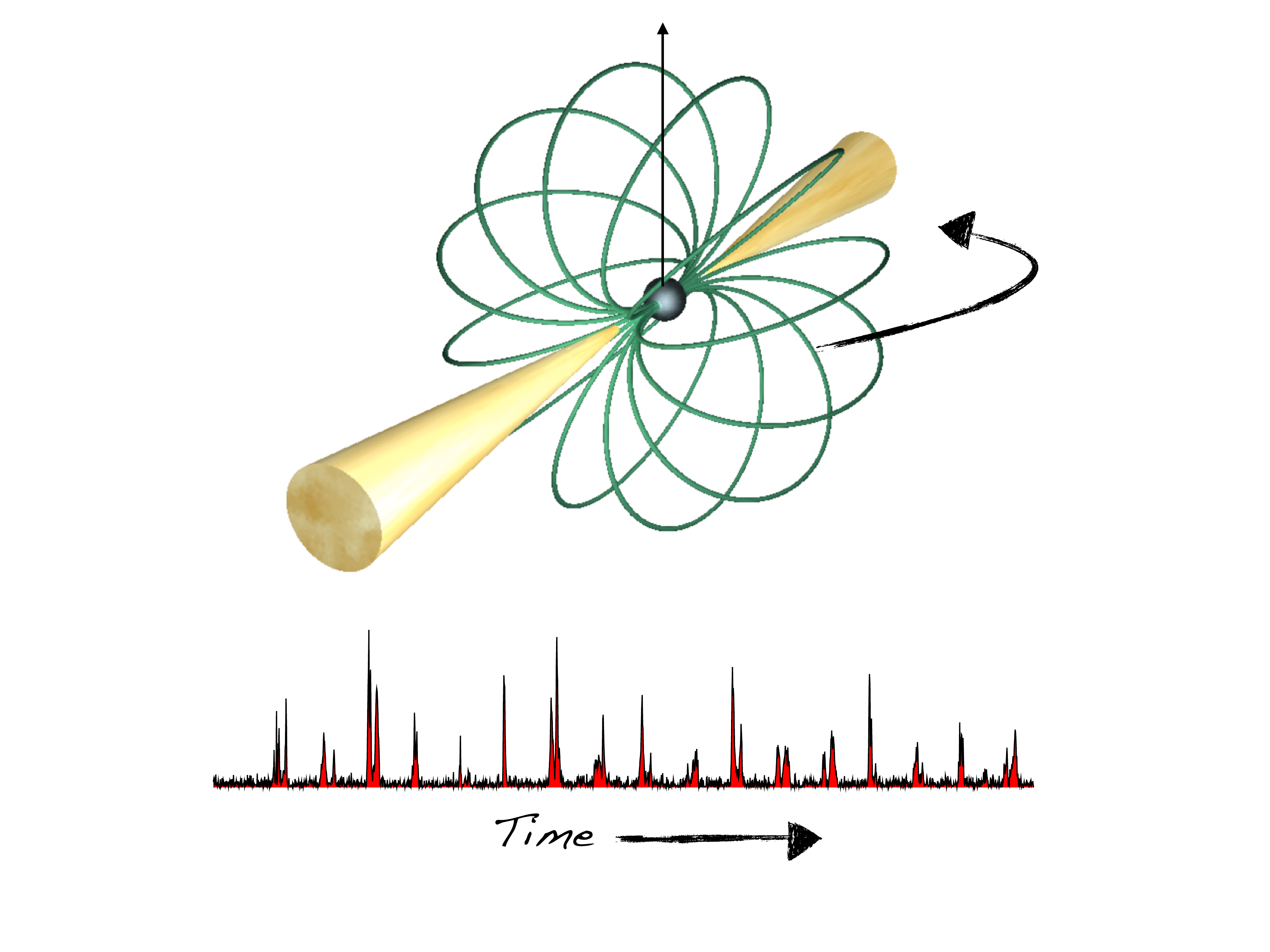}}
  \caption{\small Artist's impression of a rotation-powered 
    pulsar. The magnetized neutron star appears as a pulsating source 
    of radiation if the rotating emission beam crosses the observer's 
    line of sight. Averaging these periodic pulses of intensity over 
    many rotation cycles results in a stable pulse profile. Because of 
    the timing stability of most pulsars, the arrival time of pulses 
    can be predicted with very high precision, which is an essential 
    requirement for all applications based on pulsar 
    timing (Fig. by M.~Kramer).} \label{image:RotPowPSR}
  \end{figure}

\subsection{Pulsars}

Pulsars are born in supernova explosions of massive stars, created in
the collapse of the progenitors core.  Unlike most other
astrophysical objects, pulsars emit across the whole electromagnetic
spectrum (from radio to optical, X- and gamma-rays) at the expense of
their rotational energy, i.e., the pulsar spins down as rotational
energy is radiated away by its co-rotating magnetic field, a plasma
wind, and broadband electromagnetic radiation. Thereby all radiation
is powered by rotational energy, which distinguishes pulsars from
``accretion powered'' neutron stars.  With the magnetic axis being
inclined to the rotation axis, the pulsar acts like a cosmic
light-house emitting a radio pulse that can be detected once per
rotation when the beam is directed towards Earth (cf.~Fig.~\ref{image:RotPowPSR}).

\begin{figure}[h]
\centerline{\includegraphics[width=10.5cm]{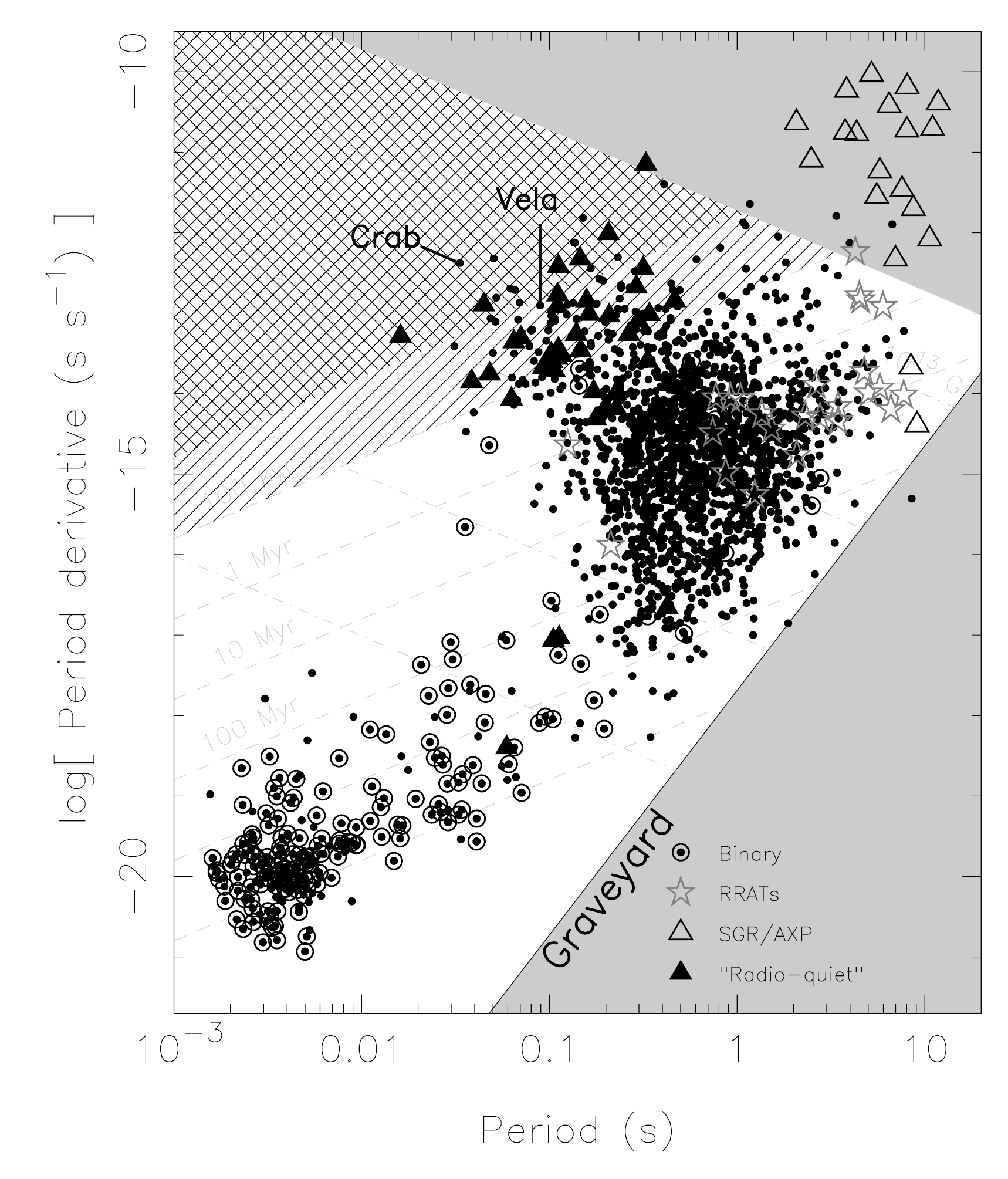}}
\caption[]{The $P-\dot{P}$--diagram for the known pulsar 
  population. Characteristic age ($\tau = P/2\dot{P}$), surface magnetic 
field ($B=3.2\times 10^{19} \sqrt{P\dot{P}} $ G) 
and spin-down luminosity ($\dot{E}= 4\pi I\dot{P}/P^3$, with $I$
being the moment of inertia) are functions of $P$ and 
$\dot{P}$ and are hence indicated as lines of corresponding 
value. Binary pulsars are marked by 
a circle. The lower solid line represents the pulsar ``death line'' 
enclosing the ``pulsar graveyard'' where pulsars are expected to 
switch off radio emission. The gray area in the top right corner 
indicates the region where the surface magnetic field appears to 
exceed the quantum critical field of $4.4\times 10^{13}$ Gauss. For 
such values, some theories expect the quenching of radio emission in 
order to explain the radio-quiet ``magnetars'' (i.e.~Soft-gamma ray 
repeaters, SGRs, and Anomalous X-ray pulsars, AXPs).}
\label{fig:ppdot}
\end{figure}

The loss in rotational energy leads to an increase in rotation period,
$P$, described by a measured $\dot{P}>0$.  Equating the corresponding
energy output of the dipole to the loss rate in rotational energy, we
obtain an estimate for the magnetic field strength at the pulsar
surface. Typical values are of order $10^{12}$~G, although field
strengths up to $10^{14}$~G have been observed \cite{msk+03}.
Millisecond pulsars (MSPs) have lower field strengths of the order of $10^8$
to $10^{10}$~G which appear to be a result of their evolutionary
history. The evolution can be tracked by two parameters, the observed
rotational slow-down, $\dot{P}$, and the resulting evolution in pulsar
period, $P$. This is usually done in a (logarithmic)
$P$-$\dot{P}$-diagram as shown in Figure~\ref{fig:ppdot}.

Most known pulsars have spin periods between 0.1 and 1.0 s with period
derivatives of typically $\dot{P}=10^{-15}$ s s$^{-1}$.  Selection
effects are only partly responsible for the limited number of pulsars
known with very long periods, the longest known period being 8.5 s
 \cite{ymj99}.  The dominant effect is due to the ``death'' of pulsars
when their slow-down has reached a critical state. This state seems to
depend on a combination of $P$ and $\dot{P}$ known as the {\em pulsar
death-line}. The normal life of radio pulsars is limited to a few tens
or hundreds million years or so.

The described evolution does not explain the over 200 pulsars in the
lower left of the $P-\dot{P}$-diagram (Fig.~\ref{fig:ppdot}). These pulsars have
simultaneously small periods (few milliseconds) and small period
derivatives ($\dot{P}\le 10^{-18}$ s s$^{-1}$).  These {\em
  millisecond pulsars} (MSPs) are much older than ordinary pulsars
with ages up to $\sim 10^{10}$ yr.  MSPs evolve from pulsars with a
binary companion. Once the binary companion evolves and overflows its
Roche lobe, it transfers mass and thereby angular momentum
(e.g.~\cite{acrs82}).  In this process, previously ``dead'' pulsars are
recycled to MSPs via an accreting X-ray binary phase.  This has a
number of observational consequences: a) most normal pulsars do not
develop into a MSP as they have long lost a possible companion during
their violent birth event; b) for surviving binary systems, X-ray
binary pulsars represent the progenitor systems for MSPs; c) the final
spin period of recycled pulsars depends on the mass of the initial
binary companion. A more massive companion evolves faster, limiting
the duration of the accretion process; d) the majority of MSPs have
low-mass white-dwarf companions as the remnant of the binary
star. These systems evolve from low-mass X-ray binary systems; e)
high-mass X-ray binary systems represent the progenitors for double
neutron star systems (DNSs). DNSs are rare since these systems need to
survive a second supernova explosion. The resulting MSP is only mildly
recycled with a period of tens of millisecond. This picture explains
the observation that $\sim 80$\% of all MSPs are in a binary orbit
while this is true for only less than 1\% of the non-recycled
population.  For MSPs with a low-mass white dwarf companion the orbit
is nearly circular. In case of DNSs, the orbit is affected by the
unpredictable nature of the kick imparted onto the newly born neutron
star in the asymmetric supernova explosion of the companion. If the
system survives, the result is typically an eccentric orbit with an
orbital period of a few hours. However, there is also evidence for the
existence of low-kick supernova, producing DNSs with low eccentricity
and relatively low-mass neutron star companions (e.g.~\cite{tlp15}).

 Source with the largest estimated magnetic fields ($\sim 10^{15}$ G),
 the so called {\em magnetars}, are located in the upper right corner 
 of Fig.~\ref{fig:ppdot}. Here, the observed luminosity appears to    
 exceed the neutron stars' spin-down energy loss, suggesting that 
 magnetars in addition to the spin-down energy are powered by 
 converting magnetic field energy (see e.g.~\cite{kk16} for a comparison 
 of magnetars to rotation powered radio pulsars). Only four magnetars
 have been detected as (transient) radio sources while all appear to 
 be X-ray and gamma-ray sources. The long-term timing is not regular, 
 so that applications as discussed below, are unlikely to be possible.

From timing measurements of binary MSPs (see Section\ref{sec:timing}), 
we can measure neutron star masses. Those are found in a range between 
about 1.2 and 2$M_\odot$ \cite{of16,ato+16} with the the maximum mass 
\cite{afw+13} ruling out the softest equation-of-sate \cite{dpr+10}. 
The MSP mass distribution is strongly asymmetric. The diversity in spin 
and orbital properties of high-mass NSs suggests that this is most 
likely not a result of the recycling process, but rather reflects 
differences in the NS birth masses. The asymmetry is best accounted 
for by a bimodal distribution with a low mass component centered at 
$1.39M_\odot$ and a high-mass component with a mean of $1.81M_\odot$ 
\cite{ato+16}.  These equations-of-state yield radii not too different 
from the very first calculations by Oppenheimer \& Volkov \cite{ov39}, 
i.e.~about 20 km in diameter, and are consistent with the blackbody 
emission radii determined from X-ray observations \cite{of16}. This 
makes neutron stars the most compact objects in the observable Universe.

\subsection{Pulsar Timing: pulsars as clocks} \label{sec:timing}

\begin{figure}[h]
 \centerline{\includegraphics[width=8cm]{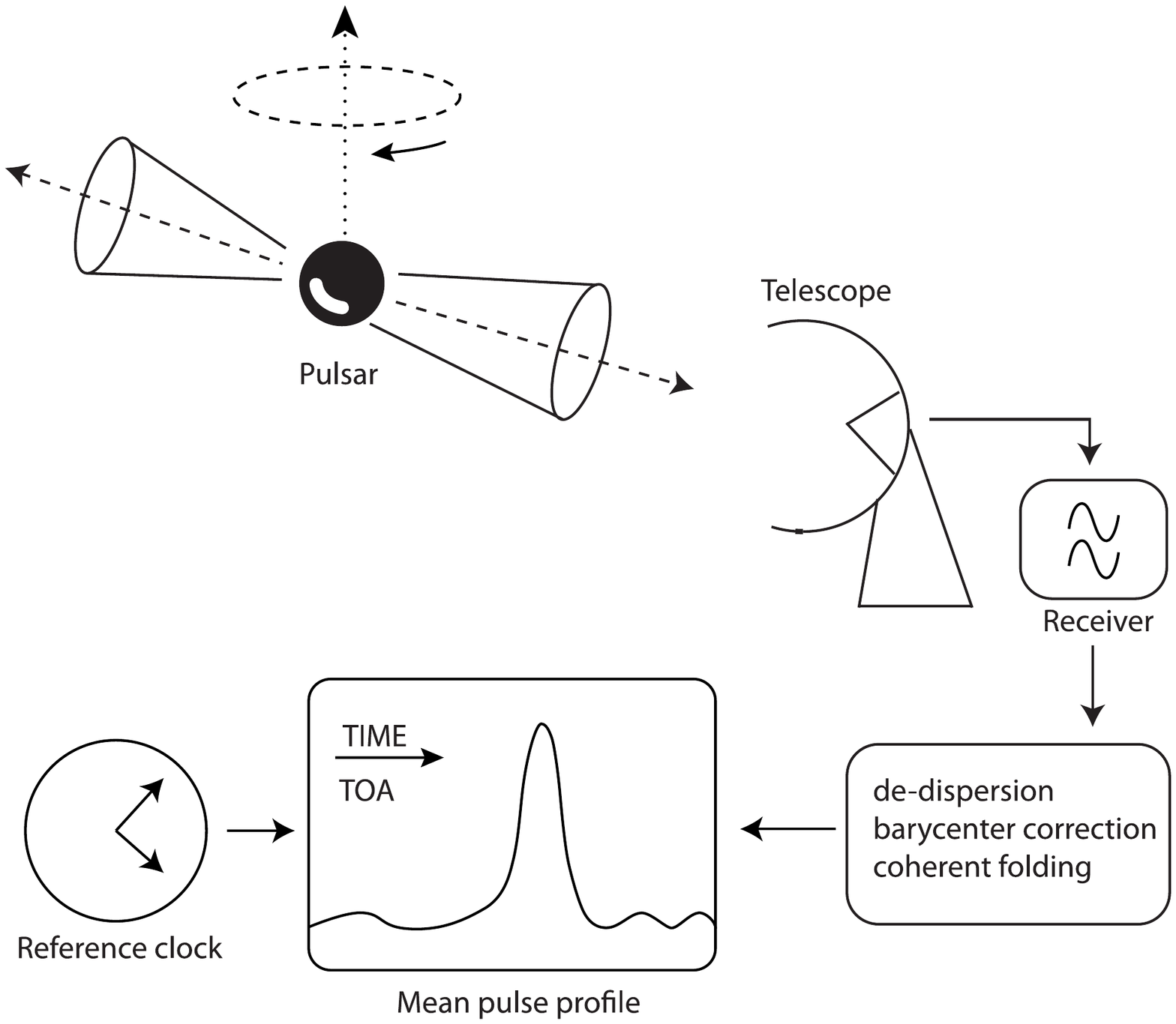}}
 \caption{Typical pulsar detection chain. The pulsar beams sweep across the radio antenna.
  Radio signals are recorded and analyzed in order to produce a mean pulse profile. The data 
  processing comprises a removal of dispersion effects caused by the interstellar medium     
  (``de-dispersion''), correction for the position and proper motion of the observatory       
  (``barycenter correction'') and coherent folding of many  pulses. The time of arrival (TOA) 
  of the pulse peak is measured against a reference
  clock. (Fig.~adapted from D.~Lorimer)}\label{RadioDetectionChain}
 \end{figure}

The basic principle of pulsar timing is the measurement of a
``time-of-arrival'' (TOA) of pulses and their identification with a specific 
rotation number of the neutron star (cf.~Fig.~\ref{RadioDetectionChain}). 
The aim is to obtain a ``coherent timing solution'', where the term ``coherent'' refers to a
complete description of the rotational phase. The experiment is repeated
many times, and the measured TOAs are compared to the prediction of the
timing solution. Deviations measured as timing residuals are minimized
by an adjustment of the timing parameters. Eventually, the
uncertainties in the rotational model are so small, that not only no
rotation is missed or mislabeled between observations, but such that
the timing model is able to predict the arrival time to microsecond
precision or better for observations decades into the future.

Radio pulsars are usually too weak to be detected in their single pulses, so 
that first an average pulse must be formed to increase the signal-to-noise 
ratio. Single pulses also usually differ in shape, intensity and exact pulse 
phase (see Fig.~\ref{image:RotPowPSR}), but occur within a well-defined window
given by the average pulse shape. Therefore, using an average pulse
also improves the timing precision and allows the usage of a technique
known as template matching. With the average pulse shape expected to
be constant between observations, one can compare the measured pulse
form with a high signal-to-noise ``template'' obtained from the
addition of many earlier observations by using a cross-correlation.
Assuming that the discretely sampled profile, ${\cal P}(t)$, is a
scaled and shifted version of the template, ${\cal T}(t)$, with added
noise, ${\cal N}(t)$, we may write
\begin{equation}
\label{equ:matching}
{\cal P}(t) = a + b{\cal T}(t-\tau) + {\cal N}(t)
\end{equation}
where $a$ is an arbitrary offset and $b$ a scaling factor.  The time
shift between the profile and the template, $\tau$, yields the TOA
relative to a fiducial point of the template and the start time of the
observation \cite{tay92}. Hence, the TOA is thereby defined as the
arrival time of the nearest pulse to the mid-point of the observation.
The uncertainty of a TOA measurement, $\sigma_{\rm TOA}$ improves with
signal-to-noise ratio (hence, size of the used telescope) and
sharpness of the pulse features, as this enables a more precise
cross-correlation result. For MSPs, a few thousand
pulses can be added easily in a few minutes of observing time. This
usually results in extremely stable profiles.  In addition to their
higher rotational stability and short duration pulses, this represents
an important factor in explaining the superior timing stability of
MSPs when compared to normal pulsars.

By measuring the arrival time of the pulsar signals very
precisely, we can study effects that determine the propagation of the
pulses in four-dimensional space-time. As indicated, the aim is
to determine the rotation number of an observed
pulse, counting from some reference epoch, $t_0$.  We can write
\begin{equation}
\label{spindown}
N = N_0 + \nu_0\times (t-t_0) + \frac{1}{2}\dot{\nu}_0\times (t-t_0)^2 + ...,
\end{equation}
where $N_0$ is the pulse number and 
$\nu_0$ the spin frequency at the reference time, respectively.
Whilst for most MSPs a second derivative, $\ddot{\nu}$,
is usually too small to be measured, we expect $\nu$ and 
$\dot{\nu}$ to be related via the physics of the
braking process, 
\begin{equation}
\label{equ:nudot}
\dot{\nu} = - \mbox{const.} \times \nu^{n}.
\end{equation}
For magnetic dipole braking  the {\em braking index} takes the value $n=3$.  If $\nu$ and its
derivatives are accurately known and if $t_0$ coincides with the
arrival of a pulse, all following pulses should appear at integer
values of $N$ --- when observed in an inertial reference
frame. However, our observing frame is not inertial, as we are using
telescopes that are located on a rotating Earth orbiting the
Sun. Therefore, we need to transfer the pulse times-of-arrival (TOAs)
measured with the observatory clock ({\em topocentric arrival times})
to the center of mass of the solar system as the best approximation to
an inertial frame available. The transformation of a topocentric TOA
to such {\em barycentric arrival times}, $t_{\rm SSB}$, is given by
\begin{eqnarray}
t_{\rm SSB} = &  & t_{\rm topo} - t_0 + 
  t_{\rm corr} - k\times {\rm DM}/f_{\rm obs}^2 \; ,
 \label{transA} \\
      & + & \Delta_{{\rm Roemer,} \odot} + \Delta_{{\rm Shapiro,}
 \odot} + \Delta_{{\rm Einstein,} \odot} \; , \label{transB} \\
      & + & \Delta_{\rm Roemer, Bin} +  \Delta_{\rm Shapiro, Bin} +
  \Delta_{\rm Einstein, Bin}. \label{transC} 
\end{eqnarray}
where DM is the so called dispersion measure, representing the
integrated path length of free electron along the line of sight,
and $f_{\rm obs}$ is the observing radio frequency (see below).
We have split the transformation into three lines. The first two lines
apply to every pulsar whilst the third line is only applicable to
binary pulsars. 

\subsubsection{Clock and frequency corrections}

The observatory time is typically maintained by local Hydrogen-maser
clocks monitored by GPS signals. In a process involving a number of
steps, clock corrections, $t_{\rm corr}$, are retroactively applied to
the arrival times in order to transfer them to a uniform atomic time
that would be kept by an ideal atomic clock on
the geoid. It is published retroactively by the {\em Bureau
International des Poids et Mesures} (BIPM).

The free electrons in the interstellar medium interact with the
propagation radio signal, causing a frequency dependent group
velocity. Consequently, the frequency components of the broadband
pulse signal emitted a higher radio frequency $f_{\rm obs}$ arrives earlier than
the corresponding low frequency components. Hence, due to this
dispersion, the measured arrival time depends on the observing
frequency, $f_{\rm obs}$ and the dispersion measure (DM).  The TOA is therefore
corrected for a pulse arrival at an infinitely high frequency (last
term in Eqn.~\ref{transA}).  For the best pulsars, the limiting factor
in timing precision is often ``interstellar weather'' that causes
small changes in DM and hence time-varying drifts in the TOAs. In
those cases, the above term is, for instance, extended to include
time-derivatives of DM which can be measured using multi-frequency
observations.

\subsubsection{Barycentric corrections}\label{barycentercorr}

The {\em Roemer delay}, $\Delta_{{\rm Roemer,} \odot}$, is the
classical light-travel time between the phase center of the telescope
and the solar system barycenter (SSB). Given a unit vector,
$\hat{s}$, pointing from the SSB to the position of the pulsar
and the vector connecting the SSB to the observatory, $\vec{r}$, we
find:
\begin{equation}
\label{roemer}
\Delta_{{\rm Roemer,} \odot}
= - \frac{1}{c}\; \vec{r} \cdot \hat{s}
= - \frac{1}{c}\left(
\vec{r}_{\rm SSB} + \vec{r}_{\rm EO} \right) 
\cdot {\hat{s}}.
\end{equation}
Here $c$ is the speed of light and we have split $\vec{r}$ into two
parts.  The vector $\vec{r}_{\rm SSB}$ points from the SSB to the center
of the Earth (geocenter). Computation of this vector requires accurate
knowledge of the locations of all major bodies in the solar system and
uses solar system ephemerides. The second vector $\vec{r}_{\rm
EO}$, connects the geocenter with the phase center of the
telescope. In order to compute this vector accurately, the non-uniform
rotation of the Earth has to be taken into account, so that the
correct relative position of the observatory is derived. 

The {\em Shapiro delay}, $\Delta_{{\rm Shapiro,} \odot}$, is a
relativistic correction that corrects for an extra delay due to the
curvature of space-time in the solar system \cite{sha64}. It is
largest for a signal passing the Sun's limb ($\sim 120$ $\mu$s) while
Jupiter can contribute as much as 200 ns. In principle one has to sum
over all bodies in the solar system, but in practice only the Sun is
usually taken into account.

The last term in Eqn.~\ref{transB}, $\Delta_{{\rm Einstein,}
\odot}$, is called {\em Einstein delay} and it describes the combined
effect of gravitational redshift and time dilation due to motion of
the Earth and other bodies, taking into account the variation of an
atomic clock on Earth in the varying gravitational potential
as it follows its elliptical orbit around the
Sun  \cite{bh86}.

\subsubsection{Relative motion \& Shklovskii effect}
\label{shlovskii}

If the pulsar is moving relative to the SSB, the transverse component
of the velocity, $v_t$, can be measured as the vector $\hat{s}$ in
Eqn.~(\ref{roemer}) changes with time.  Present day timing precision is
not sufficient to measure a radial motion although it is theoretically
possible. This leaves Doppler corrections to observed periods, masses
etc.~undetermined. The situation changes if the pulsar has an
optically detectable companion such as a white dwarf for which Doppler
shifts can be measured from optical spectra.

Another effect arising from a transverse motion is the {\em Shklovskii
effect}, also known in classical astronomy as {\em secular
acceleration}. With the pulsar motion, the projected distance of the
pulsar to the SSB is increasing, leading to 
an increase in any observed change of periodicity, such as pulsar
spin-down or orbital decay. The observed pulse period derivative, for
instance, is increased over the intrinsic value by
\begin{equation}
\label{eqn:sch}
\left(\frac{\dot{P}}{P}\right)_{\rm obs} = 
\left(\frac{\dot{P}}{P}\right)_{\rm int}
+ \frac{1}{c}\;\frac{v_t^2}{d}.
\end{equation}
For MSPs where $\dot{P}_{\rm int}$ is small, a significant
fraction of the observed change in period can be due to the
Shklovskii effect.

\begin{table}[h!!!!!!!!!!!!!!!!!!!!!!!]
\caption{Constraining specific (classes of) gravity theories using radio pulsars.
See Refs.~\cite{wex14,kk16} for details.}
{
\begin{tabular}{lp{7cm}@{\hspace{1em}}l}
\hline\noalign{\smallskip}
Theory (class)  & Method & Ref. \\
\noalign{\medskip}
\hline
\noalign{\medskip}  
\underline{Scalar-tensor gravity:} & & \\
\noalign{\smallskip}  
Jordan-Fierz-Brans-Dicke &  limits by PSR J1738+0333 and PSR
J0348+0432, comparable to best Solar system test (Cassini)  & \cite{fwe+12} \\
\noalign{\smallskip}  
Quadratic scalar-tensor gravity &  for $\beta_0 < -3$ and $\beta_0 >
0$ best limits from PSR-WD systems, in particular PSR J1738+0333 and
PSR J0348+0432 &  \cite{fwe+12}  \\
\noalign{\smallskip}   
Massive Brans-Dicke & for $m_\varphi \sim 10^{-16}$ eV: PSR J1141$-$6545 &  \cite{abwz12}\\
\noalign{\medskip}  
\hline  
\noalign{\medskip}   
\underline{Vector-tensor gravity:} & & \\
\noalign{\smallskip}    
Einstein-\AE{}ther & combination of pulsars (PSR J1141$-$6545, 
PSR J0348+0432, PSR J0737$-$3039, PSR J1738+0333) &  \cite{ybby14}\\
\noalign{\smallskip}  
Ho{\v r}ava gravity & combination of pulsars (see above) & \cite{ybby14}   \\
\noalign{\medskip}  
\hline 
\noalign{\medskip}  
\underline{TeVeS and TeVeS-like theories:} & & \\
\noalign{\smallskip}    
Bekenstein’s TeVeS & excluded using Double Pulsar  &  \cite{ks+14} \\
\noalign{\smallskip}   
TeVeS-like theories & excluded using PSR 1738+0333 & \cite{fwe+12} \\
\noalign{\medskip}
\hline
\end{tabular}}\label{tab:theories}
\end{table}

\subsubsection{Binary pulsars}

Equation~(\ref{transA}-\ref{transB}) is used to transfer the measured TOAs to
the SSB. If the pulsar has a binary companion, the light-travel time across
the orbit and further relativistic effects need to be taken into account (see
Eqn.~\ref{transC}). That adds additional orbital parameters to the set of
timing parameters which have to be solved in the timing process (see below).
In the simplest case, five Keplerian parameters need to be determined, 
i.e.~orbital period, $P_{\rm b}$; the projected
semi-major axis of the orbit, $x\equiv a \sin i$ where $i$ is the (usually
unknown) orbital inclination angle; the orbital eccentricity, $e$; the
longitude of periastron, $\omega$; and and the time of periastron passage,
$T_0$. For a number of binary systems this Keplerian description of the orbit
is not sufficient and corrections need to be applied. These can be either time
derivatives of Keplerian parameters or parameters describing completely new
effects (e.g.~those of a Shapiro delay due to curved spacetime near the
companion). In any case, it is important to note that we do not have to assume
a particular theory of gravity when measuring such relativistic corrections,
called ``post-Keplerian'' (PK) parameters  \cite{dd85,dd86}. Instead, we can
take the observational values and compare them with predictions made within a
framework of specific theories of gravity  \cite{dt92}.

 In GR, the five most important PK parameters are given
by the expressions below \cite{rob38,bt76,dd86,pet64}.

\begin{eqnarray}
\dot{\omega} &=& 3 T_\odot^{2/3} \left( \frac{P_{\rm b}}{2\pi} \right)^{-5/3} \;
               \frac{1}{1-e^2} \; (M_p + M_c)^{2/3}, \label{equ:omegadot}\\
\gamma  &=& T_\odot^{2/3} \left( \frac{P_{\rm b}}{2\pi} \right)^{1/3} \;
              e\frac{M_c(M_p+2M_c)}{(M_p+M_c)^{4/3}}, \\
r &=& T_\odot M_c, \\
s &=& \sin i =
   T_\odot^{-1/3} \left( \frac{P_{\rm b}}{2\pi} \right)^{-2/3} \; x \;
              \frac{(M_p+M_c)^{2/3}}{M_c}, \\
\dot{P}_{\rm b} &=& -\frac{192\pi}{5} T_\odot^{5/3}  \left( \frac{P_{\rm b}}{2\pi} \right)^{-5/3} 
               f(e)
               \frac{M_p M_c}{(M_p + M_c)^{1/3}}, \label{equ:pbdot}
\end{eqnarray}
where all masses are expressed in solar units, $G$ is 
Newton's gravitational constant, $c$ the speed of light and
\begin{equation}
\label{equ:fe}
f(e) = \frac{\left(1+(73/24)e^2+(37/96)e^4\right)}{(1-e^2)^{7/2}}.
\end{equation}
$P_b$ is the period and $e$ the eccentricity of the binary
orbit. The masses $M_p$ and $M_c$ of pulsar and companion,
respectively, are expressed in solar masses ($M_\odot$) where we
define the constant $T_\odot=GM_\odot/c^3=4.925490947 \mu$s. $G$
denotes the Newtonian constant of gravity and $c$ the speed of
light.

The first PK parameter, $\dot{\omega}$, describes the relativistic advance of
periastron in rad~s$^{-1}$.  It is the easiest to measure for orbits with
non-zero eccentricities (note that $\omega$ is only poorly defined for
$e\approx 0$ and so is $\dot{\omega}$). From a measurement of $\dot{\omega}$,
we obtain from Equation~(\ref{equ:omegadot}) the total mass of the system,
$(M_p+M_c)$.

The orbital decay due to gravitational-wave damping is expressed by the
(dimensionless) change in orbital period, $\dot{P}_{\rm b}$.  Any metric
theory of gravity that embodies Lorentz-invariance in its field equations
predicts gravitational radiation and, hence, $\dot{P}_{\rm b}$.  If a theory
satisfies the strong equivalence principle, like GR,
gravitational {\em dipole} radiation is not expected, but {\em quadrupole}
emission will be the lowest multipole term.  In alternative theories, while
the {\em inertial} dipole moment may remain uniform, the {\em gravitational
  wave} dipole moment may not, and 
dipole radiation may be predicted. The magnitude of this effect depends on the
difference in gravitational binding energies, expressed by the difference in
coupling constants to a scalar gravitational field. 

The other two parameters, $r$ and $s$, are related to the Shapiro delay caused
by the curvature of space-time due to the gravitational field of the
companion.  They are measurable, depending on timing precision, if the orbit
is seen nearly edge-on.

\subsubsection{Obtaining a timing solution}

Equations~(\ref{spindown}) to (\ref{transC}) contain the set of timing 
parameters that need to be determined. Many of them are
not known a priori (or only with limited precision after the discovery
of a pulsar) and need to be determined precisely in a least squares
fit analysis of the measured TOAs. The parameters can be categorized
into three groups: (a) {\em astrometric parameters} (i.e.~position,
proper motion, parallax 
contained in the R\"omer and Shapiro delay, respectively); (b) {\em
spin parameters} (i.e.~rotation frequency, $\nu$, and higher
derivatives); (c) {\em binary
parameters}.

Given a minimal set of starting parameters, a least squares fit is
needed to match the measured arrival times to pulse numbers according
to Equation~(\ref{spindown}). We minimize the expression
\begin{equation}
\chi^2=\sum_i\left( \frac{N(t_i)-n_i}{\sigma_i}\right)^2
\end{equation}
where $n_i$ is the nearest integer to $N(t_i)$ and $\sigma_i$ is the TOA
uncertainty in units of pulse period (turns). 

In order to obtain a phase-coherent solution that accounts for every
single rotation of the pulsar between two observations, one starts off
with a small set of TOAs that were obtained sufficiently close in time
so that the accumulated uncertainties in the starting parameters do
not exceed one pulse period.  Gradually, the data set is expanded,
maintaining coherence in phase.  When successful, post-fit residuals
expressed in pulse phase show a Gaussian distribution around zero with
a root mean square that is comparable to the TOA
uncertainties. Incorrect or incomplete timing models cause systematic
structures in the post-fit residuals identifying the parameter that
needs to be included or adjusted. The precision of the parameters
improves with length of the data span and the frequency of
observation, but also with orbital coverage in the case of binary
pulsars.

\begin{table}[h!!!!!!!!!!!!!!!!]
\caption{Examples of obtained relative precision using pulsar timing.
Corresponding (example) references are   cited in the last column.}
\begin{center}
{\begin{tabular}{l@{\hspace{1em}}l@{\hspace{1em}}l}
\hline\noalign{\smallskip}
Clock properties: & & \\
Spin frequency	&	$\klesssim 10^{-15}$   & \cite{vbv+08}  \\
Orbital period		 & $\klesssim 10^{-11}$ &     \cite{ks+14}   \\ 
Frequency stability & $\klesssim 10^{-15}$   & \cite{hcm+12}  \\
\noalign{\medskip}   
Masses measurements: & & \\
Neutron star & $\klesssim 10^{-4}$   &  \cite{wnt10} \\ 
White dwarf companion   & $\klesssim 10^{-3}$  & \cite{hbo06} \\
Main sequence star companion	&	$\klesssim 10^{-2}$ & \cite{fbw+11}\\
Mass of Jupiter and moons              & $\klesssim 10^{-6}$ &    \cite{chm+10} \\
\noalign{\medskip} 
Astrometry:& & \\
Distance      &         	 		$\klesssim 10^{-2}$	& \cite{vbv+08}\\
Proper motion: &  		    	$\klesssim 10^{-15}$ & 	\cite{vbv+08}\\
\noalign{\medskip} 
Gravity tests: & & \\
Test of General Relativity 	&	$\klesssim 10^{-4}$ & \cite{ks+14} \\
Constancy of grav. constant, $\dot{G}/G$ &           $\klesssim 10^{-12}$  &             \cite{fwe+12}\\
GW chara. strain $\Delta L/L$ & $\klesssim 10^{-16}$ & \cite{2016MNRAS.458.1267V} \\
\noalign{\smallskip}
\hline 
\end{tabular}}\label{tab:precision}
\end{center}
\end{table}

\subsection{Pulsars as tools}

Pulsar combine a number of interesting properties: they are compact
objects with strong gravitational fields; they are small enough to
essentially act as test masses in binary systems; the act as natural
precision clocks; their interior contains the most extreme dense
matter in the observable Universe; they emit coherent radio emission
that is up to 100\% polarised; they are not only radio sources but
pulsar emission is often seen across the whole electromagnetic
spectrum.  There are many more properties that one could list, but
just those are sufficient to make pulsars an exciting and useful tool
for probing a wide range of physics or for probing the interstellar or
other surrounding media.  They are especially useful for testing
theories of gravity. This can be done either with pulsars as part of a
binary system, or also as part of a Galactic-sized detector for
low-frequency gravitational waves. We will expand on the latter in
more detail below, but concentrate for a moment on binary pulsars.

While, strictly speaking, binary pulsars move in the weak
gravitational field of a companion, they do provide precision tests of
the (quasi-stationary) strong-field regime. This becomes clear when
considering that the majority of alternative theories predicts strong
self-field effects which would clearly affect the pulsars' orbital
motion. Hence, tracing their fall in a gravitational potential, we can
search for tiny deviations from general relativity, providing us with
unique precision strong-field tests of gravity. For instance, in
binary systems a wide range of relativistic effects can be observed,
identified and studied (e.g.~\cite{wil14}), including concepts and
principles deeply embedded in theoretical frameworks.  If a specific
alternative theory is developed sufficiently well, one can also use
radio pulsars to test the consistency of this
theory. Table~\ref{tab:theories} summarizes some, where this has been
possible. In particular, the radiative properties of a theory are a
very powerful and sensitive probe, so that every successful theories
has to pass the binary pulsar experiments.

The precision of pulsars allow also applications that may not be
directly obvious. One such application is space navigation that we
will describe in detail below. Key is the clock-like
stability of pulsars \cite{taylor1991,matsakis1997}, allowing us
to make extremely precise measurements. Table~\ref{tab:precision}
gives an idea about the precision of pulsar timing that we achieve 
today. With larger telescopes being built (e.g.~the South African MeerKAT),
precision and the number applications will increase. Perhaps, at some point, 
we will establish a complete pulsar-based timescale.

\section{Autonomous Spacecraft Navigation Based On Pulsar Timing}
\label{sec:2}

 Possible implementations of autonomous navigation systems were already discussed
 in the early days of space flight \cite{battin1964}. In principle, the orbit of a
 spacecraft can be determined by measuring angles between solar system bodies and  
 astronomical objects; e.g., the angles between the Sun and two distant stars and a
 third angle between the Sun and a planet. However, because of the limited angular 
 resolution of on-board star trackers and sun sensors, this method yields spacecraft
 positions with uncertainties that accumulate typically to several thousand kilometers.
 Alternatively, the navigation fix can be established by observing multiple solar 
 system bodies: It is possible to autonomously triangulate the spacecraft position
 from images of asteroids taken against a background field of distant stars. This   
 method was realized and flight-tested on NASA's Deep-Space-1 mission between       
 October 1998 and December 2001. The Autonomous Optical Navigation (AutoNav) system 
 on-board Deep Space~1 provided the spacecraft orbit with $1\sigma$ errors of 
 $\pm 250$\,km and $\pm 0.2$\,m/s, respectively \cite{riedel2000}. 

 An alternative and very appealing approach to autonomous spacecraft navigation
 is based on pulsar timing. The idea of using these celestial sources as a natural 
 aid to navigation goes back to the 1970s when Downs \cite{downs1974} investigated the 
 idea of using pulsating radio sources for interplanetary navigation. He analyzed a 
 method of position determination by comparing TOAs at the spacecraft 
 with those at a reference location. Within the limitations of technology and pulsar 
 data available at that time (a set of only 27 radio pulsars were considered), Downs 
 showed that spacecraft position errors on the order of 1500\,km could be obtained 
 after 24 hours of signal integration.  A possible improvement in precision by a 
 factor of 10 was estimated if better (high-gain) radio antennas were available 
 for the observations. 
 
 Chester \& Butman \cite{chester+butman1981} adopted this idea and proposed to use 
 X-ray pulsars, of which about one dozen were known at the time. They estimated
 that 24 hours of data collection from a small on-board X-ray detector with
 0.1\,m$^2$ collecting area would yield a three-dimensional position accurate to
 about 150\,km. Their analysis, though, was not based on detailed simulations or 
 actual pulsar timing analyses; neither did it take into account the technological 
 requirements or weight and power constraints for implementing such a navigation system. 

 These early studies on pulsar-based navigation estimated relatively
 large position and velocity errors so that this method was not
 considered to be an applicable alternative to the standard navigation
 schemes. However, pulsar astronomy has improved considerably over the
 last 40 years since these early proposals. Meanwhile, pulsars have
 been detected across the electromagnetic spectrum and their emission
 properties have been studied in great detail (see
 Section~\ref{intro}). Along with the recent advances in detector and
 telescope technology this motivates a general reconsideration of the
 idea of pulsar-based spacecraft navigation.

 \subsection{The relevance of the various pulsar types for navigation\label{TypesOfPulsars}\label{subsec:2}}

 We already discussed the various types of pulsars, namely rotation-,
 accretion-, and magnetic-field-powered neutron stars in
 Section~\ref{intro}.  The very complex spin behavior and often 
 unpredictable evolution of rotation period in accretion powered
 pulsars manifest themselves in erratic changes between spin-up and
 spin-down as well as X-ray burst activities.  This unsteady and
 non-coherent timing behavior disqualifies them as reference sources
 for navigation. Similar arguments for magnetars invalidates these
 sources also for the use in a pulsar-based spacecraft navigation
 system.

 Concerning their application for navigation, the only pulsar class
 that really qualifies is that of rotation-powered ones. Here, the
 much higher timing stability of MSPs is of major
 importance for their use in a pulsar-based navigation system.
Of the $\sim2500$ rotation-powered pulsars known today
 (Fig.~\ref{fig:ppdot}), about 150 have been detected in the X-ray
 band \cite{becker2009}, and approximately $1/3$ of them are MSPs.
 In the past $30-40$ years many of them have
 been regularly timed with high precision especially in radio
 observations. Consequently, their ephemerides (RA, DEC, $P$,
 $\dot{P}$, binary orbit parameters, TOAs and absolute
 pulse phase for a given epoch, pulsar proper motion etc.) are known
 with very high accuracy (see Table~\ref{tab:precision}).  This is an
 essential requirement for using these celestial objects as navigation
 beacons, as it enables us to predict the TOAs of a
 pulsar for any location in the solar system and beyond.

  \begin{figure}[h]
  \centerline{\includegraphics[width=8cm, angle=90,clip=]{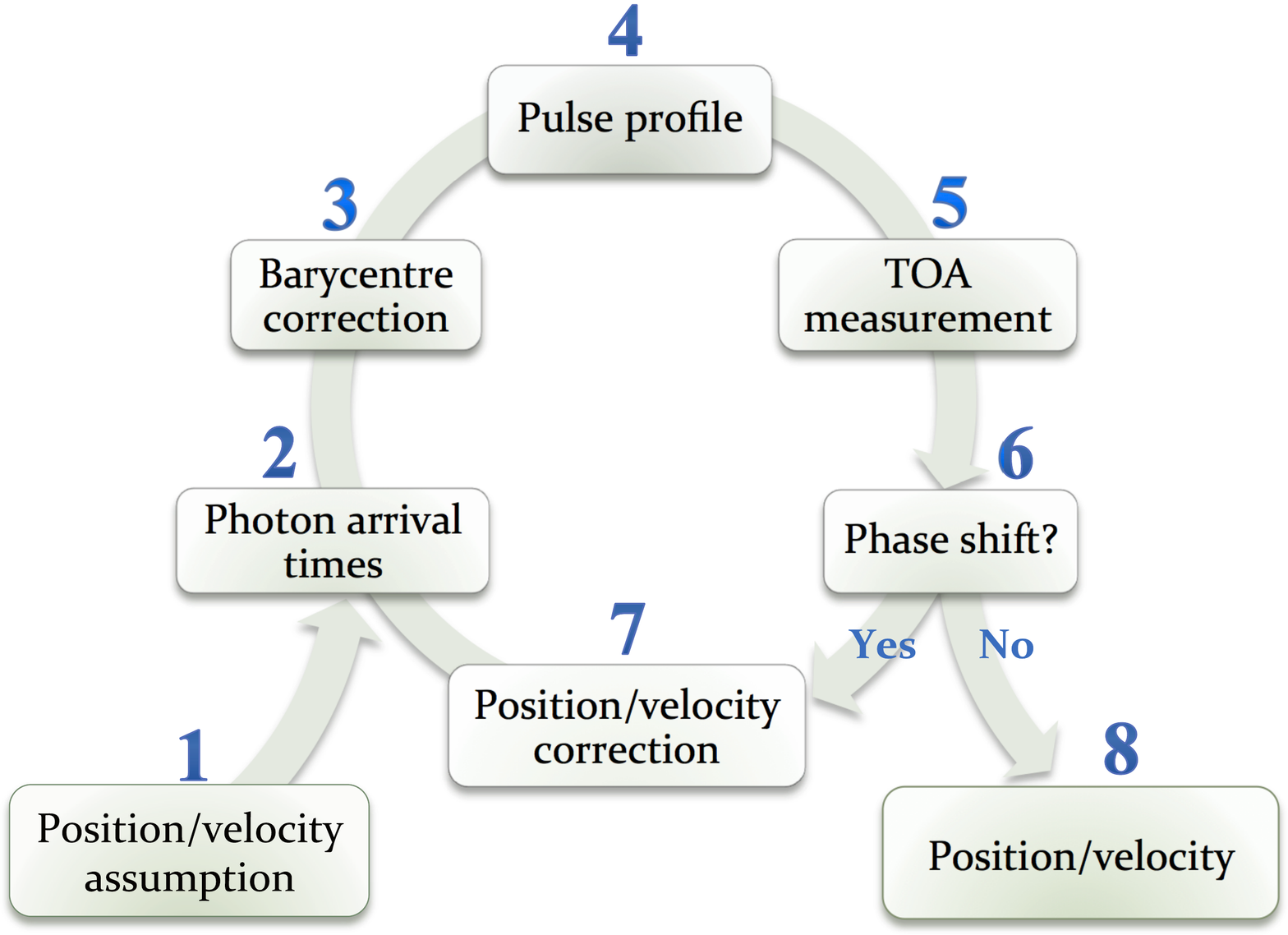}}
  \caption{\small Iterative determination of position and velocity by a pulsar-based 
  navigation system \cite{2015AN....336..749B}.}\label{image:iteration_loop}
  \end{figure}

\subsection{Timing irregularities}

So far, we have neglected that young rotation-powered pulsars can
show {\em glitches} in their spin-down behavior, i.e., abrupt
increases of rotation frequency, often followed by an exponential
relaxation toward the pre-glitch frequency \cite{espinoza2011,
  yu2013}. While this is often observed in young pulsars, it is very rarely in
old and millisecond pulsars. Nevertheless, the glitch behavior of
pulsars has to be taken into account by a pulsar-based navigation
system. While it is possible to change the set of pulsars taken as
reference when a glitch has been observed, the intrinsic {\em timing
  noise} is a factor which sets a hard limit for the overall accuracy
possible for a pulsar-based spacecraft navigator. Timing noise is
present in all pulsars, but the level varies between sources. Typical MSPs
show timing residuals with an root mean square (RMS) 
at a level of $\sim 0.1-1 \mu$s, which limits the
position accuracy then to 30 - 300m.

 \begin{figure}[h]
 \centerline{\includegraphics[width=8cm, angle=0,clip=]{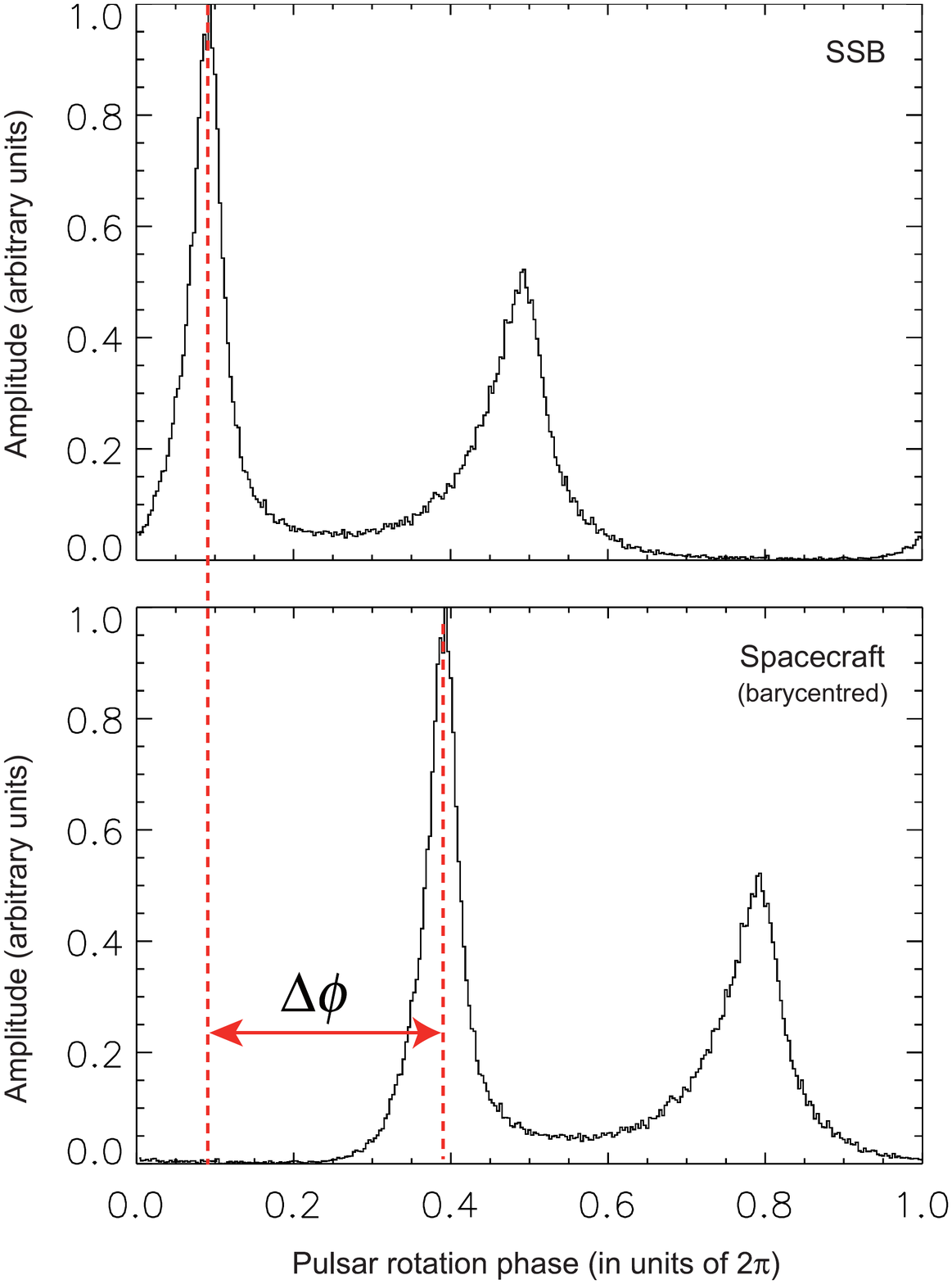}}
 \caption{\small Measuring the phase difference between the expected 
  and measured pulse peak at an inertial reference location; e.g., the 
  solar system barycenter (SSB). The top profile shows the main pulse peak 
  location as expected at the SSB. The bottom profile is the one measured 
  at a spacecraft and transformed to the SSB by {\rm assuming} the spacecraft 
  position and velocity during the observation. If the position and  velocity 
  assumption was wrong, a phase shift $\Delta \phi$ is observed \cite{2015AN....336..749B}.}
  \label{image:Crab_pulseprofile}
 \end{figure}

\subsection{Principles of Pulsar-Based Navigation\label{PrinciplesPulsarNavigation}}
\label{sec:xraytiming}

The concept of using pulsars as navigational aids is based on the
comparison of measured TOA with predicted TOA at a given epoch and
reference location. Figure~\ref{RadioDetectionChain} showed the
typical chain for detecting radio signals from a rotation-powered
pulsars. For our application here, we perform x-ray observations,
where the dedispersion in not necessary. However, the step of applying
barycenter correction to the observed photon arrival times (see
Section~\ref{barycentercorr}) is still essential.  The pulsar
ephemerides along with the position and velocity of the observer are
parameters of this correction. Using a spacecraft position that
deviates from the true position during the observation results in a
phase shift of the pulse peak (or equivalently in a difference in the
pulse arrival time). Therefore, the position and velocity of the
spacecraft can be adjusted in an iterative process until the pulse
arrival time matches with the expected one. The corresponding
iteration chain is shown in Figure~\ref{image:iteration_loop}.

 \begin{figure}[h]
 \centerline{\includegraphics[width=9cm, angle=0,clip=0]{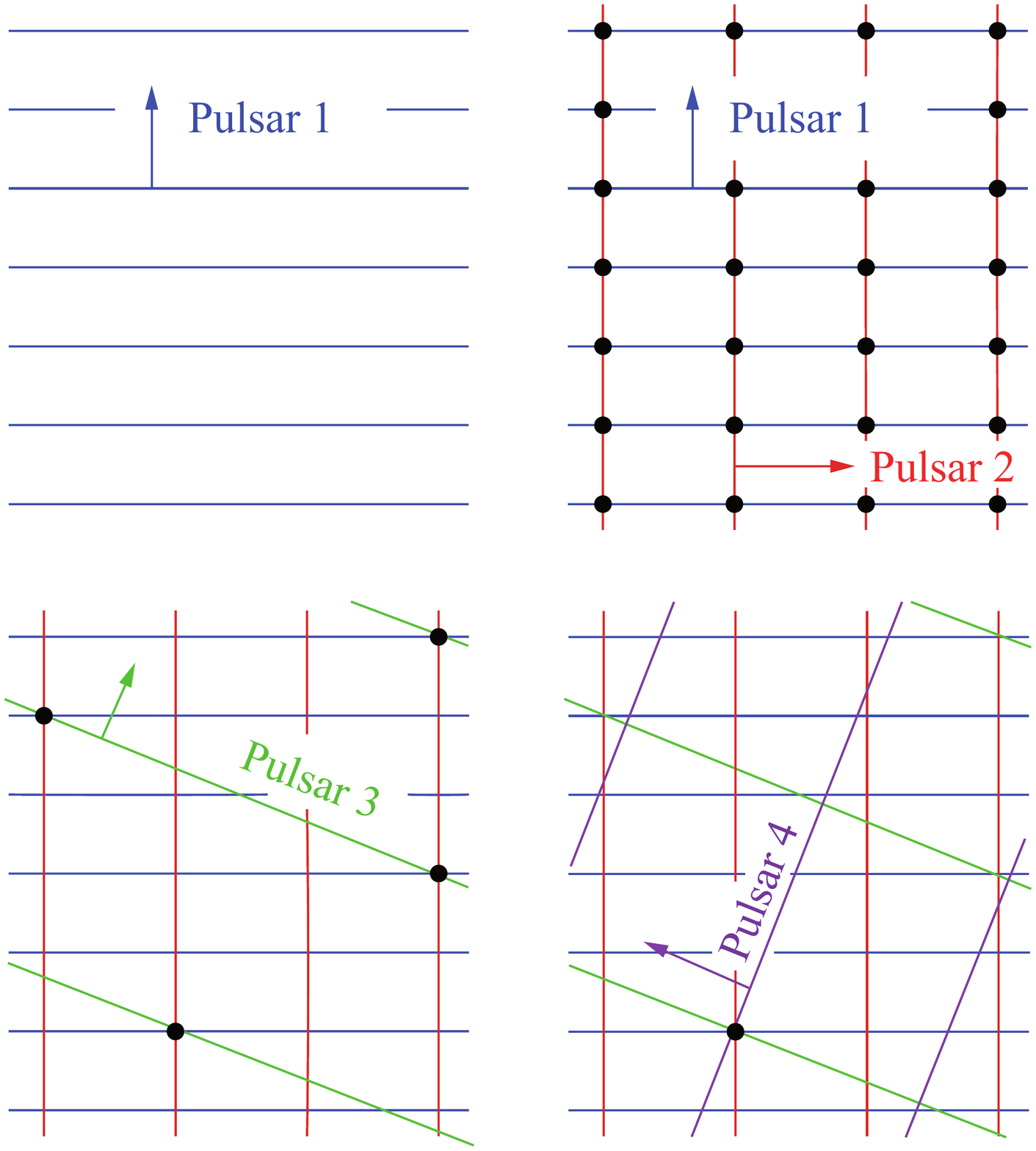}}
 \caption{\small Solving the ambiguity problem by observing four pulsars (drawn in two dimensions). 
  The arrows point along the pulsar's lines-of-sight. Straight lines represent planes of constant  
  pulse phase; black dots indicate intersections of planes \cite{2015AN....336..749B}.}\label{image:three_pulsars}            
 \end{figure} 

 An initial assumption of position and velocity is given by the
 planned orbital parameters of the spacecraft (1). The iteration
 starts with a pulsar observation, during which the arrival time of 
 individual photons are recorded (2). The photon arrival times have to be
 corrected for the proper motion of the spacecraft by transforming the
 arrival times to an inertial reference location; e.g., the solar
 system barycenter (SSB). As for pulsar observations with
 terrestrial telescopes, this correction requires knowledge of the
 (assumed or deduced) spacecraft position and velocity as input
 parameters. The barycenter corrected photon arrival times allow then
 the construction of a pulse profile or pulse phase histogram (4)
 representing the temporal emission characteristics and timing
 signature of the pulsar.  This pulse profile, which is continuously
 improving in significance during an observation, is permanently
 correlated with a pulse profile template in order to increase the
 accuracy of the absolute pulse-phase measurement (see
 Section~\ref{sec:timing}) (5), or equivalently, TOA.  From the pulsar ephemeris that includes the information of
 the absolute pulse phase for a given epoch, the phase difference
 $\Delta\phi$ between the measured and predicted pulse phase can be
 determined (cf.~Fig.~\ref{image:Crab_pulseprofile}).

 In this scheme, a phase shift (6) with respect to the absolute pulse phase corresponds to a
 range difference $\Delta x = c P (\Delta\phi + n)$ along the line of sight toward the observed
 pulsar. Here, $\Delta\phi$ the phase shift and
 $n=0, \pm 1, \pm 2,\ldots$ an integer that takes into account the periodicity of the observed pulses.
 If the phase 
 shift is non-zero, the position and velocity of the spacecraft needs to be corrected accordingly
 and the next iteration step is taken (7). If the phase shift is zero, or falls below a certain  
 threshold, the position and velocity used during the barycenter correction was correct (8) and  
 corresponds to the actual orbit of the spacecraft.

 A three-dimensional position fix can be derived from observations of at least three
 different pulsars (cf.~Figure~\ref{image:three_pulsars}). If on-board clock calibration
 is necessary, the observation of a fourth pulsar is required.

 Since the position of the spacecraft is deduced from the phase (or TOA)
 of a periodic signal, ambiguous solutions may occur. This problem can be solved by constraining
 the domain of possible solutions to a finite volume around an initial assumed position 
 \cite{bernhardt2010, bernhardt2011}, or by observing  additional pulsars as illustrated
 in Figure~\ref{image:three_pulsars}.

\section{Gravitational Wave Detection Based On Pulsar Timing}
\label{sec:3}

\subsection{Principles of gravitational wave detection with pulsar timing}

Because of their exquisite stability, MSPs also constitute a formidable tool for 
the detection of low frequency gravitational waves (GWs). The underlying concept 
is very simple: GWs affect the propagation of radio signals from the pulsar to the 
Earth, leaving a characteristic signature in the pulses TOAs. If this effect is not 
included in the timing solutions described in Section \ref{sec:timing}, it will 
show up in pulsar ``$a$'' as a residual of the form \cite{sv10}:

\begin{equation}
  r_a(t) = \int_0^{t} \frac{\delta \nu_a}{\nu_a}(t') dt',
\label{res}
  \end{equation}
where
\begin{equation}
\frac{\delta \nu_a}{\nu_a} = \frac{1}{2} \frac{\hat{p}_a^i \hat{p}_a^j}{1 + \hat{p}_a.\hat{\Omega}} \Delta h_{ij}.
\label{dnuovernu}
\end{equation}
Here $\nu_a$ is the MSP spin frequency, $\hat{p}_a$ denotes the position of 
the pulsar on the sky, and $\hat{\Omega}$ is the direction of the GW propagation. The quantity

\begin{equation} \label{metric}
  \Delta h_{ij} = h_{ij}(t_a^p) - h_{ij}(t)
\end{equation}
is the strain of the GW at the location of the pulsar $h_{ij}(t^p_a)$ and on Earth $h_{ij}(t)$ 
(indices $i,j$ run on the three spatial coordinates), and the pulsar time $t_a^p$ is related 
to the Earth time $t$ as:

\begin{equation}
  t_a^p = t - L_{a}(1 + \hat{\Omega}.\hat{p}_{a}) \equiv t - \tau_{a},
\end{equation}
where $L_{a}$ is the distance to the pulsar. In practice, GWs cause
drifts in the MSP spin frequency 
(redshift/blueshift) proportional to the difference of the perturbing field at the pulsar and 
on Earth. Those drifts cause the pulse to arrive at the Earth earlier/later than expected, 
resulting in the distinctive timing residual $r_a(t)$ of equation (\ref{res}). The properties 
of $r_a(t)$ ultimately depend on the nature of the perturbation $\Delta h_{ij}$ which can be 
either a deterministic function or a stochastic process. Both cases are astrophysically relevant, 
and analysis methods have been developed accordingly.

Pulsars are complex astrophysical objects, governed by largely unknown physics that might affect 
their intrinsic stability and the GW signal has to be disentangled from a plethora of other effects 
(see \cite{2015MNRAS.453.2576L} for a comprehensive discussion). For example, a deterministic 
sinusoidal GW signal might be virtually indistinguishable from the effect of a binary companion. 
Therefore, the GW residual signature needs to be simultaneously detected in several MSPs. This is 
even more crucial if the GW signal has a stochastic nature. In this case, $r_a(t)$ is just a 
realization of a stochastic process, at par with other stochastic noise sources. GW detection 
therefore requires the cross correlation of the TOAs from an ensemble of MSPs, forming a pulsar 
timing array (PTA, \cite{1990ApJ...361..300F}). Three main PTA collaborations are now active in 
the monitoring of several tens of MSPs: The European Pulsar Timing Array (EPTA \cite{2016MNRAS.458.3341D}), 
the Parkes Pulsar Timing Array (PPTA \cite{2016MNRAS.455.1751R}) and the North American Nanohertz 
Observatory for Gravitational Waves (NANOGrav, \cite{2015ApJ...813...65T}). The three collaborations
are constantly improving their data, that they also share since 2010 under the aegis of the 
International Pulsar Timing Array (IPTA, \cite{2016MNRAS.458.1267V}), aiming at the formation of a 
combined, more sensitive dataset. 

Altogether, the three PTAs are timing order of fifty of the best MSPs at a weekly cadence ($\Delta{t}$) 
for a timespan $T$ of several years (more then 20 for some cases), with a timing precision of few 
micro-seconds to few tens of nano-seconds. PTAs are therefore sensitive to GWs in the frequency 
range $1/T<f<1/(2\Delta{t})$, corresponding to few to few-hundred nano-Hertz. Possible GW sources 
in this frequency range include cosmological stochastic backgrounds from inflation, phase transitions 
or cosmic strings \cite{2016PhRvX...6a1035L}, but the loudest signals are expected from the cosmic 
population of inspiralling supermassive black hole binaries (SMBHBs) formed following galaxy mergers \cite{svc08}.

\subsection{Signals from supermassive black hole binaries}
\label{subsec:3.1}

In the astrophysically reasonable assumption of circular, monochromatic, non-precessing binary, orbiting 
at a frequency $f$, the two independent polarization amplitudes generated by the system can be written 
as \cite{sv10}:

\begin{subequations}
\begin{align}
h_+(t) & = (1 + \cos^2 \iota) A_\mathrm{gw} \cos\Phi(t)\,,
\\
h_{\times}(t) &=-2 \cos\iota \, A_\mathrm{gw} \sin\Phi(t)\,,
\label{e:hpluscross}
\end{align}
\end{subequations}
where
\begin{equation}
A_\mathrm{gw}(f) = 2 \frac{{\cal M}^{5/3}}{D}\,\left[\pi f(t)\right]^{2/3}
\label{e:Agw}
\end{equation}
is the GW amplitude, $D$ the luminosity distance to the GW source, 
${\cal M}=(M_1M_2)^{3/5}/(M_1+M_2)^{1/5}$ is the chirp mass (being $M_1$ and $M_2$ the 
masses of the two SMBHs) and $\Phi(t) = 2\pi\int^t f(t') dt'$ is the GW phase. The metric 
perturbation in equation (\ref{metric}) can therefore be written as:
\begin{equation}
h_{ij}(t,\hat\Omega) = e_{ij}^+(\hat\Omega) (h_+(t^p_a,\hat\Omega)-h_+(t,\hat\Omega)) + 
e_{ij}^{\times}(\hat\Omega)\,(h_\times(t^p_a,\hat\Omega)- h_\times(t,\hat\Omega)),
\label{e:hab}
\end{equation}
where the contribution of both the pulsar and the Earth term have been included. 
$e_{ij}^A(\hat\Omega)$ ($A = +\,,\times$) are the polarization tensors, that are uniquely defined 
once one specifies the wave principal axes described by the unit vectors $\hat{m}$ and $\hat{n}$ as,
\begin{subequations}
\begin{align}
e_{ij}^+(\hat{\Omega}) &=  \hat{m}_i \hat{m}_j - \hat{n}_i \hat{n}_j\,,
\\
e_{ij}^{\times}(\hat{\Omega}) &= \hat{m}_i \hat{n}_j + \hat{n}_i \hat{m}_j\,.
\label{e:eplusecross}
\end{align}
\end{subequations}

The signal is therefore deterministic and can be described as $r_{a}(\overrightarrow{\lambda},t)$, 
where $\overrightarrow{\lambda}$ is the vector of parameters specifying the SMBHB (including SMBH masses, 
binary sky location, inclination, frequency and initial phase). Note that $r_a$ also depends on the sky 
location of the pulsar through the response function in equation (\ref{dnuovernu}) and on the distance 
to the pulsar that affects the phase and possibly the frequency of the pulsar term 
(see \cite{2016MNRAS.455.1665B} for a full description).

Since galaxy mergers are common, we expect {\it at any time} a large population of SMBHB 
emitting GWs in the PTA band. Therefore, the superposition of many incoherent signals results 
in a stochastic GW background (GWB), with energy content described in terms of its GW energy 
density $\rho_\mathrm{gw}$ per unit logarithmic frequency, divided by the critical energy 
density, $\rho_c$, to close the Universe:

\begin{equation}
  \label{eq:omegagw}
  \Omega_{\mathrm{gw}}(f)=\frac{1}{\rho_c}\frac{\mathrm{d} \rho_{\mathrm{gw}}}{\mathrm{d}\ln f} =\frac{2\pi^2}{3H_0^2}f^2 h^2_c(f).
\end{equation}

Here, $f$ is the GW frequency, $\rho_c=3H_0^2/8\pi$ is the critical energy density required to close the 
Universe, $H_0=100~h$~km~s$^{-1}$~Mpc$^{-1}$ is the Hubble expansion rate. In the limit of circular 
GW-driven SMBHBs the `characteristic strain', $h_c(f)$, associated with a GWB energy density 
$\Omega_{\mathrm{gw}}(f)$ is parametrised as a single power-law:
\begin{equation}
h_c=A\left(\frac{f}{\mathrm{yr}^{-1}}\right)^{-2/3},
  \label{hcA}
\end{equation}
where $A$ is the strain amplitude at a characteristic frequency of 1yr$^{-1}$. Finally, $h_c$ is directly related to the 
observable quantity induced by a GWB in the timing residuals, the one-sided power spectral density, $S(f)$, given by:
\begin{equation}
  S(f) = \frac{1}{12\pi^2}\frac{1}{f^3}h_c(f)^2 = \frac{A^2}{12\pi^2}\left(\frac{f}{\rm{yr}^{-1}}\right)^{-13/3}\rm{yr}^3.
  \label{Sh}
\end{equation}

Therefore, in the case of a stochastic GWB, the signal is described by a stochastic red process with power spectral 
density given by equation (\ref{Sh}). This can be described by a vector of two parameters $\overrightarrow{\eta}=(A,\gamma)$, 
where $\gamma=-13/3$ for circular GW-driven SMBHBs. We will see below that this power has a very specific correlations 
between pairs of pulsars, which is the distinctive signature that has to be searched in PTA data.

\subsection{Current status: upper limits from pulsar timing arrays}
\label{subsec:3.3}
Whether the signal is deterministic or stochastic, the core aspect of all GW searches is the evaluation of the likelihood 
that some signal is present in the time series of the pulsar TOAs. Without entering in technical details 
(see, e.g. \cite{2013MNRAS.428.1147V,2015MNRAS.453.2576L}) the likelihood marginalized over the timing parameters 
can be written as

\begin{eqnarray}
        \mathcal{L}(\vec{\theta}, \overrightarrow{\lambda}, \overrightarrow{\eta} | \overrightarrow{\delta t}) 
        = \frac{1}{\sqrt{(2\pi)^{n-m} det(G^T C G) }} \times
\nonumber \\
\exp{\left(-\frac1{2} (\overrightarrow{\delta t} - \overrightarrow{r}(\overrightarrow{\lambda}))^T G 
(G^T C(\overrightarrow{\eta}) G)^{-1} G^T (\overrightarrow{\delta t} - \overrightarrow{r}(\overrightarrow{\lambda})) \right)}.
\label{Eq:lik}
\end{eqnarray}

Here  $n$ is the length of the vector $\overrightarrow{\delta t} = \cup \delta{t}_{a}$ obtained by concatenating the 
individual pulsar TOA series $\delta{t}_{a}$, $m$ is the total number of parameters describing the timing model 
(see Section \ref{sec:timing}), and the matrix $G$ is related to the design matrix (see \cite{2013MNRAS.428.1147V} 
for details). The variance-covariance matrix $C$, in its more general version, contains contributions from the 
GWB and from the white and (in general) red noise: $C = C_{gw}(\overrightarrow{\eta}) + C_{wn} + C_{rn}$. We refer 
the reader to \cite{2015MNRAS.453.2576L} for exact expressions of the noise variance-covariance matrix. 

Likelihoods similar to equation (\ref{Eq:lik}) have been used to search for both i) a deterministic single 
GW sources (e.g., particularly loud SMBHBs rising above the level of the stochastic GWB), or ii) a stochastic GWB.

\begin{figure}
\centerline{\includegraphics[width=14cm,angle=0,clip=1]{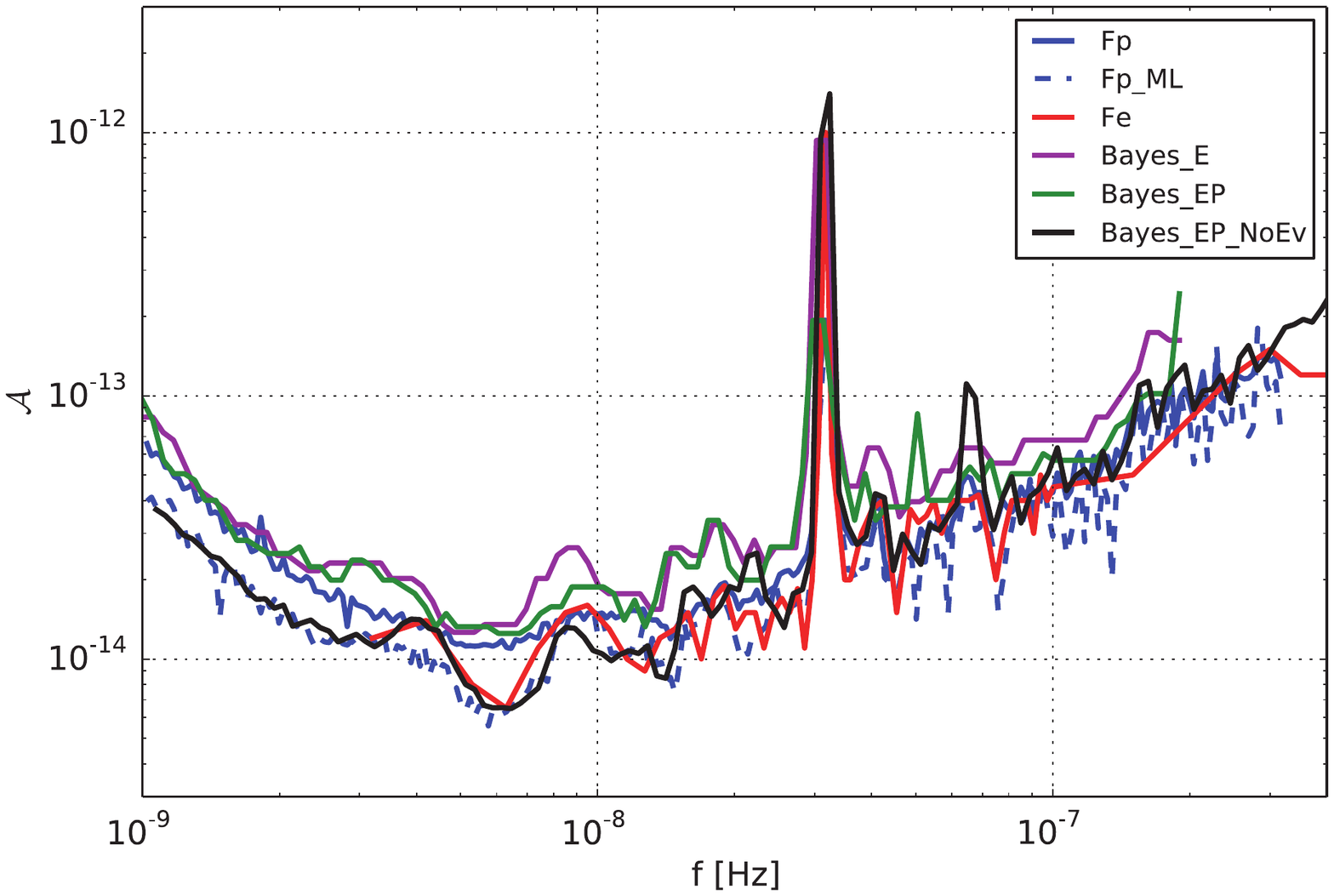}}
\caption{The 95\% upper limit on the gravitational wave strain $A_{\rm gw}$ for the 3 frequentist methods, i.e. ${\cal F}_p$ 
varying noise ({\it Fp}), ${\cal F}_p$ fixed noise ({\it Fp\_ML}) and ${\cal F}_e$, and the 3 Bayesian methods, i.e. ``evolving 
source'' with Earth term only ({\it Bayes\_E}) and with Earth and Pulsar terms ({\it Bayes\_EP}) and ``non-evolving source'' with 
Earth and Pulsar terms ({\it Bayes\_EP\_NoEv}), performed in the EPTA single source analysis. From \cite{2016MNRAS.455.1665B}, 
where the detailed descriptions of each method can be found.}
\label{fig_singleUL}
\end{figure}

In case i), a deterministic signal $r(\overrightarrow{\lambda},t)$ is added to the model and a search is performed over the 
parameter space defined by $\overrightarrow{\lambda}$. If the data are better described by noise plus a deterministic signal, 
then the properties of the latter can be inferred through the posterior distribution of $\overrightarrow{\lambda}$; otherwise, 
upper limits on the amplitude $A_{\rm gw}$ of a single GW signal (equation (\ref{e:Agw})) can be placed at every frequency. 
Searches of this type have been performed by the three major PTAs \cite{2014ApJ...794..141A,2014MNRAS.444.3709Z,2016MNRAS.455.1665B}. 
The most stringent limit to date has been placed by EPTA and is shown in figure \ref{fig_singleUL} (from \cite{2016MNRAS.455.1665B}) 
as a function of frequency for different type of searches (see \cite{2016MNRAS.455.1665B} for full details). Searches yielded 
yet no convincing evidence of individual GW sources, and the presence of SMBHBs of few billion solar masses emitting in the 
PTA range can be ruled out to the distance of the Coma cluster.

\begin{figure}
\centerline{\includegraphics[width=11.8cm,angle=0,clip=0]{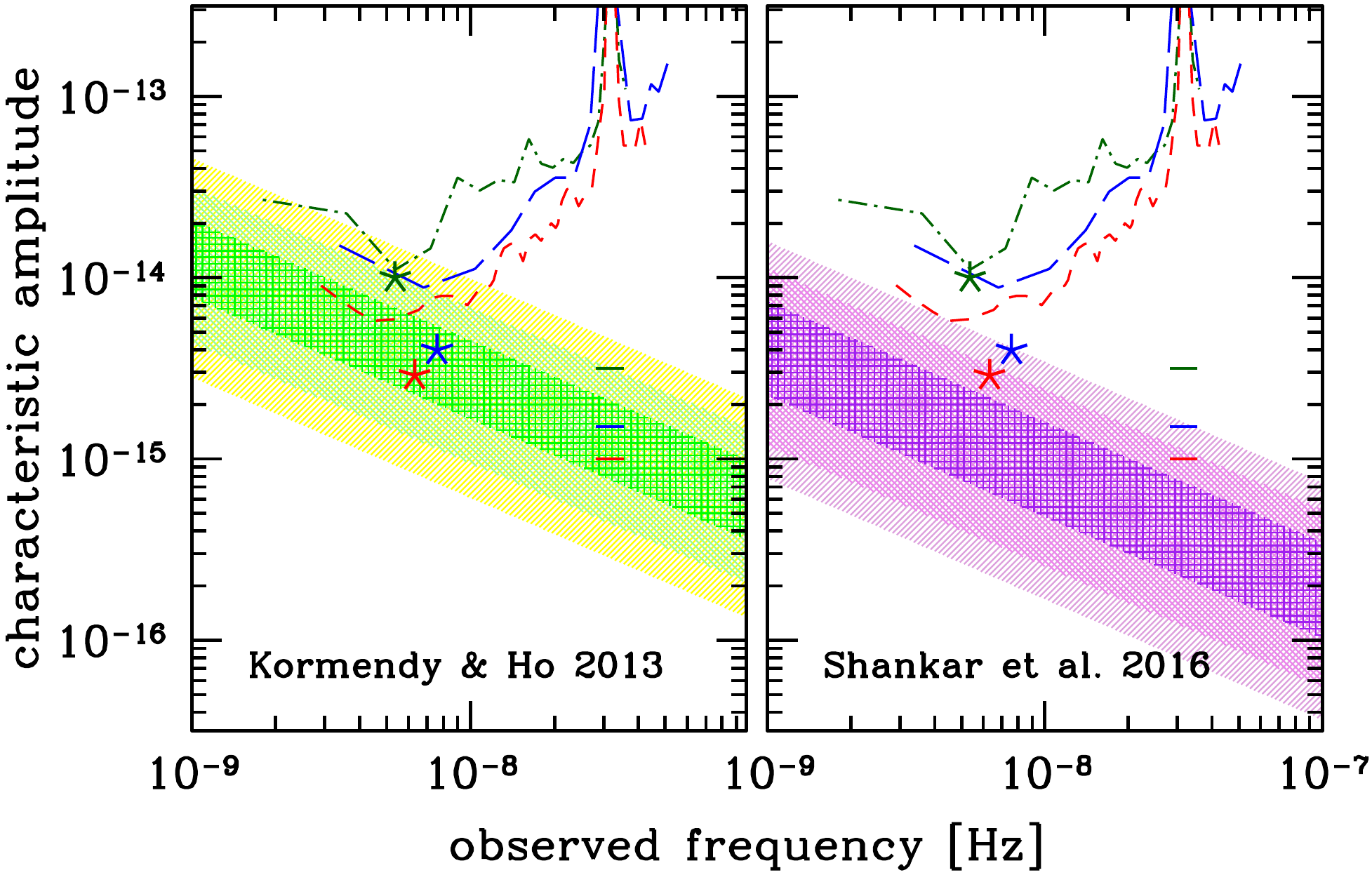}}
\caption{From \cite{2016MNRAS.463L...6S}. PTA limits on the amplitude of a stochastic GWB. The jagged curves are current 
PTA sensitivities: EPTA (dot-dashed green), NANOGrav (long-dashed blue), and PPTA (short-dashed red). For each sensitivity 
curve, stars represent the integrated upper limits to an $f^{-2/3}$ background, and the horizontal ticks are their 
extrapolation at $f=1$yr$^{-1}$, i.e. the upper limit on $A(f=1$yr$^{-1})$ quoted in the main text. The shaded areas 
represent the 68\% 95\% and 99.7\% confidence intervals of the characteristic amplitude $h_c(f)$ of the expected GWB 
form SMBHBs. The two panels represent predictions from models employing two
different $M_{\rm BH}$"-host bulge mass relations: the one from 
\cite{kh13} (left) and the one from \cite{sbs+16} (right). See \cite{2016MNRAS.463L...6S} for full 
details of the employed models.
}
\label{fig_hclimit}
\end{figure}

\begin{figure}
\centerline{\includegraphics[width=13cm]{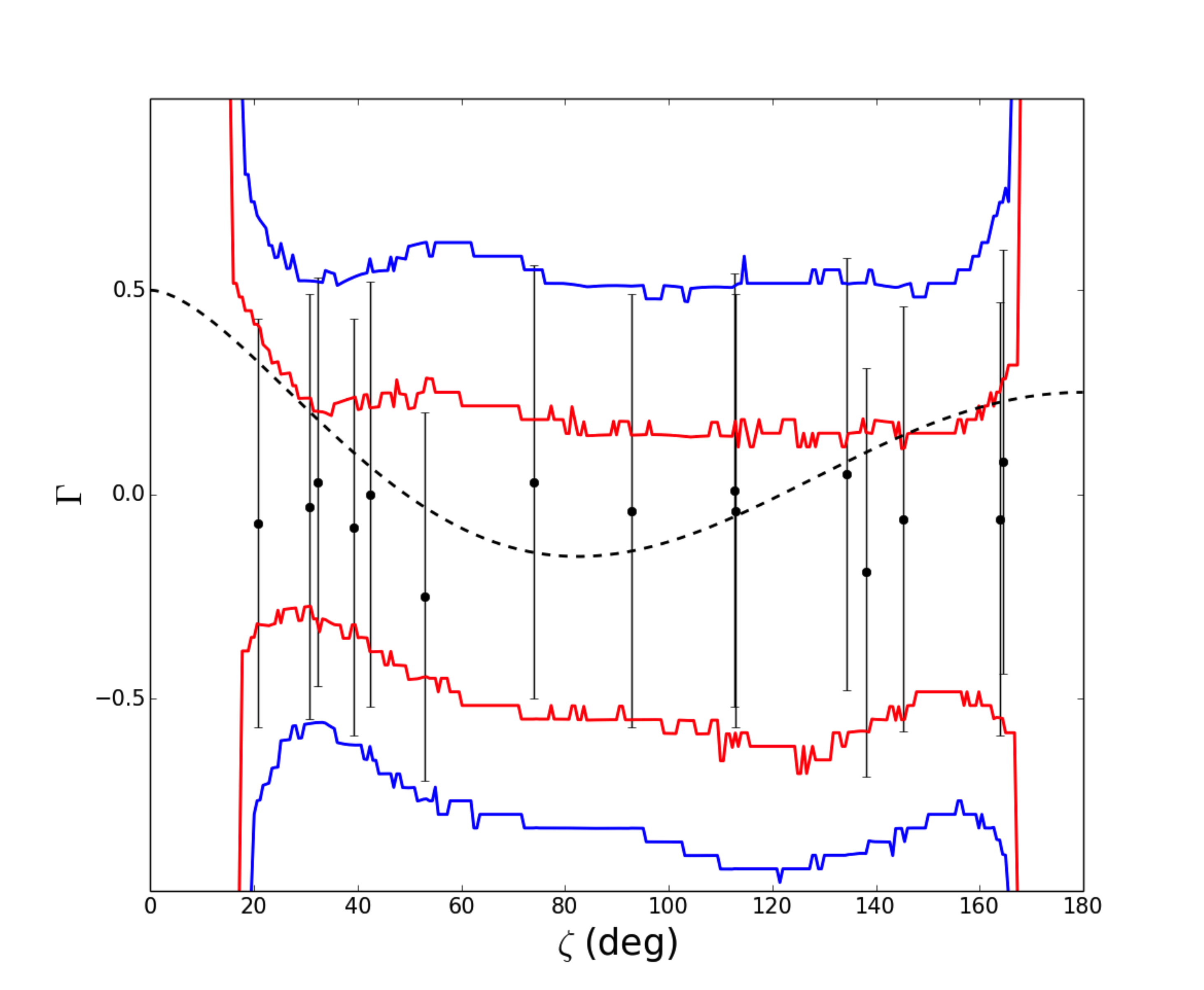}}
\caption{The recovered correlation between pulsars as a function of angular separation on the sky for a power law 
noise process in the EPTA analysis. The red and blue lines represent the $68\%$ and $95\%$ confidence intervals 
for the correlation function when modeled by the lowest 4 Chebyshev polynomials, while the individual points 
are the mean correlation coefficient with $1\sigma$ uncertainty for each pulsar pair when fitting without assuming 
a smooth model.  The Hellings-Downs correlation is represented by the dotted line. From \cite{2015MNRAS.453.2576L}.}
\label{fig_correlation}
\end{figure}

In case ii), i.e. the search for a GWB, $C_{gw}(\overrightarrow{\eta})$ is considered in the likelihood evaluation. 
The signal is stochastic in nature, and has to be singled out among other stochastic sources of noise, including red 
noise processes peculiar to individual MSPs, correlated noises due to clock or ephemeris errors etc. The smoking gun 
of a stochastic GWB is provided by the peculiar correlation pattern it introduces in the residuals of pulsar pairs 
as a function of their separation. This is a general property of any GWB described by the two tensor polarizations 
allowed by GR and was first worked out by Hellings and Downs \cite{hd83}: 

\begin{eqnarray}
\Gamma(\zeta_{IJ})=\frac{3}{8}\left[1+\frac{\cos\zeta_{IJ}}{3}+4(1-\cos\zeta_{IJ})\ln\left(\sin\frac{\zeta_{IJ}}{2}\right) \right](1+\delta_{IJ}) \, .
\label{e:hdcurve}
\end{eqnarray}

Here $\zeta_{IJ}$ is the angle between the pulsars $I$ and $J$ on the sky and $\Gamma(\zeta_{IJ})$ is the overlap 
reduction function, which represents the expected correlation between the TOAs given an isotropic stochastic GWB, 
and the $\delta_{IJ}$ term accounts for the pulsar term for the autocorrelation.

In the Fourier representation of the covariance matrix $C_{\rm GW}$, this results in a contribution 
\cite{2015MNRAS.453.2576L} 

\begin{eqnarray}
\label{Eq:FreqMatrix}
\Psi_{I,J,i,j} = \Gamma(\zeta_{IJ}) \varphi^{\mathbf{\mathrm{GWB}}}_{i}\delta_{ij},
\end{eqnarray}
that can be included in the likelihood function (\ref{Eq:lik}), appropriately evaluated in the Fourier domain. 
Lower case indices $i,j$ run over the difference frequency bins of the Fourier decomposition and $\varphi_i$ 
corresponds to the power in the GW signal given by equation (\ref{Sh}), evaluated at the central frequency 
$f_i$ if the $i$-th bin. Also in this case, one can evaluate whether the data are better described by including 
a GWB (and therefore evaluate the vector parameter $\overrightarrow{\eta}$) or not, and in the latter case 
upper limits on $S_h(f)$ (or equivalently on $h_c(f)$) can be placed as a factor of frequency. Also in this 
case, no convincing detection has been made yet. An illustrative example is given in figure \ref{fig_correlation}, 
from \cite{2015MNRAS.453.2576L}, where the measured correlation pattern in the EPTA dataset is shown as a 
function of the angular separation of the MSP pairs used in the analysis. As expected, the correlation is 
consistent with zero and the functional form (\ref{e:hdcurve}) could not be detected. With a null result, 
limits on $h_c(f)$ can be placed, as shown in figure \ref{fig_hclimit} from \cite{2016MNRAS.463L...6S}. 
Upper limits are usually quoted in term of the amplitude $A$ at $f=$yr$^{-1}$, as defined in equation 
(\ref{Sh}). Current limits are $A=3\times10^{-15},1.5\times10^{-15},10^{-15},1.7\times10^{-15}$ for 
EPTA \cite{2015MNRAS.453.2576L}, NANOGrav \cite{2016ApJ...821...13A}, PPTA \cite{2015Sci...349.1522S} 
and IPTA \cite{2016MNRAS.458.1267V} respectively. Note that, although the IPTA limit is not the most 
stringent one, it has not been obtained combining the most recent individual PTA datasets. Compared 
to the individual dataset used in the combination, the IPTA limit is a factor of $\approx2$ better, 
demonstrating the great potential of collaborating at a worldwide level. As shown in figure \ref{fig_hclimit}, 
these limits start to dig into the level expected by a cosmological population of SMBHBs, and a positive 
detection might be expected in the near future.

\section{Summary and Future Prospects}
\label{sec:4}

Pulsars are unique astrophysical sources that combine a number of
special properties, which converts them into effective cosmic
laboratories for fundamental physics. After having provided the very
first evidence for GWs \cite{tay94b}, pulsars can also be used as GW
detectors, probing a nHz-frequency range, where we expect the 
signals from supermassive binary black holes to occur.

In fact, following galaxy mergers, the SMBHs hosted at their centers
pair together forming a binary system. Since galaxy mergers are
observed to be frequent in the Universe, a large population of
adiabatically inspiralling SMBHBs is expected to fill the nHz GW
sky. Such low frequencies are inaccessible to ground and space based
GW interferometers, but are in the reach of ongoing and forthcoming
PTAs.

Current PTAs are already placing stringent limits to the maximum GW
strain emitted by these systems, starting to pierce through the level
predicted by current SMBH assembly models. As timing continues, PTA
sensitivity naturally increases and progressively extends at lower
frequencies (where the signal is expected to be the
loudest). Moreover, new instrumentation will play a pivotal role in
the coming years. MeerKAT in South Africa and FAST in China will
provide even better timing precision on a larger ensemble of pulsars,
and by the mid 20s' we will be surveying thousands of MSPs throughout
the galaxy with SKA.

With these premises, it is likely that nHz GW detection will be likely
within the next decade, placing another milestone in our understanding
of the cosmic evolution of the biggest compact objects in the
Universe.

But the clock-like properties of pulsars, and their
accessibility across the whole electromagnetic spectrum, also open the door
for a wide rage of practical applications. Here we have especially
discussed spacecraft navigation.

The knowledge of how to use stars, planets and stellar constellations
for navigation was fundamental for mankind in discovering new
continents and subduing living space in ancient times. It is
fascinating to see how history repeats itself in that a special
population of stars may play again a fundamental role in the future of
mankind by providing a reference for navigating their spaceships
through the Universe.

  Autonomous spacecraft navigating  with pulsars is feasible when using either 
  phased-array radio antennas of at least 150\,m$^2$ antenna area or compact 
  light-weighted X-ray telescopes and detectors, which are currently developed 
  for the next generation of X-ray observatories.

  Using the X-ray signals from MSPs, Becker et al.~\cite{2015AN....336..749B} estimated
  that navigation would be possible with an accuracy of $\pm 5$\,km in the solar system
  and beyond. The uncertainty is dominated by the inaccuracy of the X-ray pulse profile
  templates that were used for the pulse peak fittings and pulse-TOA measurements. As  
  those are known with much higher accuracy in the radio band, it is possible to increase the 
  accuracy of pulsar navigation down to the meter scale by using radio signals from 
  pulsars for navigation.

 %
  The disadvantage of radio observations in a navigation application, though, is the large size  
  and mass of the phased-antenna array. The antenna area is inversely proportional 
  to the square root of the integration time; i.e., the same signal quality can be 
  obtained with a reduced antenna size by increasing the observation time. However, 
  the observing time is limited by the Allen variance of the receiving system and, 
  therefore, cannot become arbitrarily large. In addition, irradiation from the 
  on-board electronics requires an efficient electromagnetic shielding to prevent 
  signal feedback. This shielding will further increase the navigator weight in 
 addition to the weight of the antenna.

  The optimal choice of the observing band depends on the boundary conditions given
  by a specific mission. What power consumption and what navigator weight might be
  allowed for may determine the choice for a specific wave band.

  In general, however, it is clear already today that this navigation technique
  will find its applications in future astronautics. The technique behind it is
  very simple and straightforward, and pulsars are available everywhere in the 
  Galaxy. Today $\approx 2500$ pulsars are known. With the next generation of radio
  observatories, like the SKA, it is expected to detect signals from about
  20\,000 to 30\,000 pulsars \cite{smits2009}.

  Finally, pulsar-based navigation systems can operate autonomously. This  
  is one of their most important advantages, and is interesting also for   
  current space technologies; e.g., as augmentation of existing GPS/Galileo
  satellites. Future applications of this autonomous navigation technique might be
  on planetary exploration missions and on manned missions to Mars or beyond.

\begin{acknowledgement}
A.S.~is supported by a University Research Fellowship of the Royal Society.
\end{acknowledgement}
%


\end{document}